\documentclass[12pt]{article}

\usepackage{amsmath,amssymb,amsfonts,graphics,graphicx,amscd,amsfonts,epsfig,color}
\usepackage{subfig}
\usepackage{here} 
\usepackage{cite}
\newcommand {\beq} {\begin{equation}}
\newcommand {\eeq} {\end{equation}}
\newcommand {\beqa}{\begin{eqnarray}}
\newcommand {\eeqa}{\end{eqnarray}}

\makeatletter
    
    \@addtoreset{equation}{section}
  \makeatother

\allowdisplaybreaks[4]

\date{}
\begin{document}

\begin{flushright} 
KEK-TH-2023
\end{flushright} 

\vspace{0.50cm}

\begin{center}
{\bf \large Complex Langevin analysis of 
the spontaneous symmetry breaking in 
dimensionally reduced super Yang-Mills models}
\end{center}

\vspace{0.5cm}

\begin{center}

         Konstantinos N. A{\sc nagnostopoulos}$^{a}$\footnote
          { E-mail address : konstant@mail.ntua.gr},  
 
         Takehiro A{\sc zuma}$^{b}$\footnote
          { E-mail address : azuma@mpg.setsunan.ac.jp},  
         Yuta I{\sc to}$^{c}$\footnote
          { E-mail address : yito@post.kek.jp},
         Jun N{\sc ishimura}$^{cd}$\footnote
          { E-mail address : jnishi@post.kek.jp} and 
          
         Stratos Kovalkov P{\sc apadoudis}$^{a}$\footnote
          { E-mail address : sp10018@central.ntua.gr}         
 
\vspace{0.5cm}
$^a$ {\it Physics Department, National Technical University,\\
Zografou Campus, GR-15780 Athens, Greece}

$^b${\it Institute for Fundamental Sciences, Setsunan University, \\
17-8 Ikeda Nakamachi, Neyagawa, Osaka, 572-8508, Japan}

$^c${\it KEK Theory Center, 
High Energy Accelerator Research Organization,\\
1-1 Oho, Tsukuba, Ibaraki 305-0801, Japan} 

$^d${\it Graduate University for Advanced Studies (SOKENDAI),\\
1-1 Oho, Tsukuba, Ibaraki 305-0801, Japan} 

\end{center}

\vspace{0.5cm}

\begin{center}
  {\bf abstract} 
\end{center}
In recent years the complex Langevin method (CLM) has proven
a powerful method in studying statistical systems which suffer
from the sign problem. 
Here we show that it can also be applied to an important problem
concerning why we live in four-dimensional spacetime.
Our target system is the type IIB matrix model, 
which is conjectured to be a nonperturbative definition of 
type IIB superstring theory in ten dimensions.
The fermion determinant of the model
becomes complex upon Euclideanization, which causes a severe sign problem
in its Monte Carlo studies.
It is speculated that the phase of the fermion determinant actually
induces the spontaneous breaking of the SO(10) rotational symmetry,
which has direct consequences on the aforementioned question.
In this paper, we apply the CLM to the 6D version of the type IIB matrix model
and show clear evidence that the SO(6) symmetry is broken down to SO(3).
Our results are consistent with those obtained previously by the
Gaussian expansion method.

\newpage

\section{Introduction}
Monte Carlo methods have been playing a crucial role in nonperturbative
studies of quantum field theories and statistical systems relevant to
particle, nuclear and condensed matter physics.
However, in many interesting cases, it happens that
such methods cannot be applied straightforwardly because
the effective ``Boltzmann weight'' can become negative or even complex.
A brute-force method would be to use the absolute value of the weight
in generating
configurations and to take into account the sign or the phase in
calculating the expectation values. This reweighting method indeed works
for small systems, but the computational cost grows exponentially
with the system size due to huge cancellations among configurations,
which is commonly referred to as the sign problem.
This problem occurs, for instance, in studying
finite density systems, including that of QCD, 
the real time dynamics of quantum systems,
supersymmetric theories and strongly correlated electron systems.

In recent years there has been major progress in evading the sign problem
by complexifying the dynamical variables, which are supposed to be real 
in the original system.
One such approach is the generalized Lefschetz-thimble 
method \cite{Cristoforetti:2012su,Cristoforetti:2013wha,Fujii:2013sra,%
Alexandru:2015sua},
which amounts to deforming the integration contour in such a way
that the sign problem becomes mild enough to be handled by the reweighting method.
Another approach is the 
complex Langevin method (CLM) \cite{Parisi:1984cs,Klauder:1983sp}, 
which attempts to
define a stochastic process
for the complexified variables so that
the expectation values with respect to this process are equal to 
the expectation values defined in the original system
extending the idea of stochastic quantization \cite{Parisi:1980ys}.
In both approaches, holomorphy plays a crucial role.
The advantage of the CLM as compared with the other is 
that it is computationally less costly, which enables its application
to much larger system size.
The disadvantage, on the other hand, is that the equivalence to the
original system does not always hold. This has been a serious issue
for a long time because one can obtain explicit 
numerical results which are simply wrong
without even noticing it.
The progress in this direction was made by clarifying the
conditions for the 
equivalence \cite{Aarts:2009uq,Aarts:2011ax,Nishimura:2015pba,%
1606_07627,Salcedo:2016kyy,Aarts:2017vrv}
and by inventing new techniques
that made it possible to meet these conditions for a 
larger space of parameters 
\cite{Seiler:2012wz,Nagata:2016alq,Nagata:2015uga,Tsutsui:2015tua,%
1609_04501,Bloch:2017ods,Doi:2017gmk,Bloch:2017sex}.


Here we aim at applying the CLM to the so-called 
type IIB matrix model \cite{hep-th9612115}, 
which is conjectured to provide a nonperturbative definition of 
superstring theory.
In its Euclidean version, this model suffers from a severe sign problem
due to the complex Pfaffian, which appears after integrating out the
fermionic matrices. The phase of the Pfaffian fluctuates 
less violently for configurations with lower 
dimensionality \cite{hep-th0003223}, 
and this effect is speculated to cause the spontaneous breaking of 
the SO(10) rotational symmetry for ten bosonic matrices that
represent the spacetime \cite{Nishimura:2000wf,hep-th0108041}.
While the original expectation was that the SO(10) symmetry is spontaneously
broken to SO(4) in order to account for the appearance of four-dimensional
spacetime \cite{Aoki:1998vn,Nishimura:2001sx}, 
explicit calculations based on the Gaussian expansion method (GEM)
suggested that it is broken down to SO(3) instead \cite{1108_1293}.
In order to address this issue by first principle calculations,
we clearly need to overcome the sign problem, which plays a crucial role in 
the phenomenon itself. 

A first step in that direction has been taken by 
two of the authors (Y.~I.\ and J.~N.) in ref.~\cite{1609_04501},
where a toy model \cite{Nishimura:2001sq} 
in which SO(4) symmetry is expected to break
spontaneously due to the complex fermion determinant
has been studied by the CLM. 
In particular, 
a new technique has been proposed
to 
avoid the so-called singular-drift 
problem \cite{Nishimura:2015pba}
caused by the eigenvalues of the Dirac operator close to 
zero \cite{Mollgaard:2013qra}.
The idea is to avoid this problem by
deforming the Dirac operator with a kind of mass term.
This makes it possible to satisfy the condition
for justifying the method 
as one can confirm by probing the
histogram of the drift term appearing in the complex Langevin 
equation \cite{1606_07627}. 
After extrapolating the deformation parameter 
to zero, the CLM was able to reproduce the
results of the GEM \cite{Nishimura:2004ts} 
\emph{including the pattern of the spontaneous symmetry breaking (SSB)},
which was not possible \cite{Anagnostopoulos:2010ux,Anagnostopoulos:2011cn}
by a Monte Carlo method \cite{hep-th0108041} based
on reweighting.

In this paper, 
we extend this work to the 6D
version of the type IIB matrix model, which can be obtained by 
dimensionally reducing 6D super Yang-Mills theory to a point
in the same way as one obtains the type IIB matrix model by
dimensionally reducing 10D super Yang-Mills theory.
The 6D version also suffers from the sign problem 
due to a complex determinant, 
which appears after integrating out the
fermionic matrices, and its phase is speculated to cause the SSB
of SO(6) rotational symmetry to SO(3) according to the GEM \cite{1007_0883}.
We show that the CLM indeed
enables us to address this issue from
first principles with the aid of the deformation technique, 
and our results turn out to be consistent with those of the GEM.
Here again, the success of the CLM is remarkable 
compared with the marginally successful results
of the reweighting-based method \cite{1306_6135}.
This gives us a big hope that 
the original type IIB matrix model
can also be studied by the CLM,
and that one can determine the pattern of the SSB,
which seemed to be extremely hard
in the reweighting-based method \cite{1509_05079}.

The rest of this paper is organized as follows.
In section \ref{sec:model} 
we briefly review the known results for dimensionally 
reduced super Yang-Mills models.
In section \ref{CLM-6dIKKT} we discuss how we apply the CLM to
the 6D version of the type IIB matrix model and present the results in
section \ref{sec:result}.
Section \ref{sec:summary} is devoted to a summary and discussions.
In appendix \ref{sec:details} we provide some details
of our complex Langevin simulation.

\section{Brief review of dimensionally reduced super Yang-Mills models} 
\label{sec:model}

As is well known, one can define
${\cal N}=1$ pure super SU($N$) Yang-Mills theories in $D=3,4,6,10$ 
dimensions.\footnote{In the $D=6$ and $D=10$ cases, the 
super Yang-Mills theories have gauge anomaly, and hence one can
consider them only at the classical level. This fact does not cause
any problem in defining the corresponding dimensionally reduced
models.}
By dimensionally reducing these theories to a point,
one obtains matrix models with $D$ bosonic matrices $A_{\mu}$ and 
their superpartners $\psi_\alpha$, 
which are called dimensionally reduced super 
Yang-Mills models in the literature.
In particular,
the $D=10$ case corresponds to
the type IIB matrix model \cite{hep-th9612115},
which has been proposed as a nonperturbative definition
of superstring theory
in the same sense as the lattice formulation 
provides a nonperturbative definition of quantum field theories.
Of particular interest is the fact that
the spacetime is
represented by 
the eigenvalue distribution of the bosonic matrices $A_{\mu}$
in this model \cite{Aoki:1998vn}, 
and hence even the spacetime dimensionality
should be determined dynamically.

In what follows, we consider the Euclidean version of the 
dimensionally reduced models, in which
the spacetime indices are contracted using the Kronecker 
delta instead of the Minkowski metric.\footnote{The Lorentzian
version of the dimensionally reduced models is also found to have
interesting dynamical 
properties \cite{Kim:2011cr,Ito:2013ywa,Ito:2015mxa,Ito:2017rcr}. 
The relationship between the two versions is not clear, though.}
The $D=3$ model is ill-defined since the partition function 
is divergent \cite{hep-th9803117,hep-th0103159}. 
The $D=4$ model has a real nonnegative fermion determinant, 
and Monte Carlo simulation suggested 
that the SO$(4)$ rotational symmetry is not spontaneously 
broken \cite{hep-th0003208}. 
The $D=6$ and 10 models have a complex fermion determinant/Pfaffian, 
whose phase is expected to play a crucial role in the SSB of 
SO$(D)$ \cite{hep-th0003223,Nishimura:2000wf,hep-th0108041,1306_6135,1509_05079}.

In this paper we therefore focus on the $D=6$ model 
defined by the partition function
\begin{eqnarray}
 Z = \int dA d \psi d {\bar \psi} \, e^{-(S_{\textrm{b}} + S_{\textrm{f}})} 
\label{IKKT_partition}
\end{eqnarray}
as a simplified model of the type IIB matrix model.
The bosonic part $S_{\textrm{b}}$ and the fermionic part $S_{\textrm{f}}$ 
of the action are given, respectively, as 
\begin{eqnarray}
 S_{\textrm{b}} &=& - \frac{1}{4} N \, 
\textrm{tr} [A_{\mu}, A_{\nu}]^{2} \ , 
\label{IKKT_boson} \\
 S_{\textrm{f}} &=& N \, \textrm{tr} 
\left( {\bar \psi}_{\alpha} (\Gamma_{\mu})_{\alpha \beta} 
[A_{\mu}, \psi_{\beta}] \right) \ .
\label{IKKT_fermion}
\end{eqnarray}
The bosonic matrices $A_{\mu}$ ($\mu =1, \cdots, 6$) are 
traceless Hermitian $N \times N$ matrices, while
the fermionic matrices
$\psi_{\alpha}$ ($\alpha =1,\cdots, 4$) are 
traceless $N \times N$ matrices with Grassmann entries. 
The action is invariant under SO(6) transformations, 
under which $A_{\mu}$ transforms as a vector,
while $\psi_{\alpha}$ transforms as a Weyl spinor.
The $4 \times 4$ gamma matrices $\Gamma_{\mu}$
obtained after the Weyl projection are given, for instance, as
\begin{eqnarray}
 \Gamma_1 &=& i \sigma_1 \otimes \sigma_2 \ , \quad
\Gamma_2  = i \sigma_2 \otimes \sigma_2 \ , \quad
\Gamma_3  = i \sigma_3 \otimes \sigma_2 \ ,  \nonumber \\
\Gamma_4  &=& i  {\bf 1} \otimes \sigma_1 \ , \quad
 \Gamma_5 = i {\bf 1} \otimes \sigma_3 \ , \quad
\Gamma_6  = {\bf 1} \otimes {\bf 1} 
\label{6d_gamma} 
\end{eqnarray}
in terms of the Pauli matrices $\sigma_i$ ($i=1,2,3$).


Integrating out the fermionic matrices $\psi_\alpha$, we obtain
\begin{eqnarray}
\int  d \psi d {\bar \psi} \,
e^{-S_{\textrm{f}}} = \det {\cal M} \ , 
\label{ferm_det}
\end{eqnarray}
where ${\cal M}$ is a $4(N^2-1) \times 4(N^2-1)$ matrix,
which represents a linear transformation 
\beq
\Psi_\alpha  \mapsto ({\cal M}\Psi)_\alpha \equiv 
 (\Gamma_{\mu})_{\alpha \beta} [A_{\mu}, \Psi_{\beta}]
\label{defM}
\eeq
acting on the linear space of 
traceless complex $N\times N$ matrices $\Psi_{\alpha}$.
The determinant $\det {\cal M}$ 
takes complex values in general, and
we define its phase $\Gamma$ by
$\displaystyle \det {\cal M} = |\det {\cal M}| \, e^{i \Gamma}$.
If one omits the phase $\Gamma$, Monte Carlo studies become straightforward,
and it is found that the SSB of SO(6) does not occur \cite{1306_6135}.
(Similar results are obtained in the 10D case as well \cite{1509_05079}.)

Let us here define ``$d$-dimensional configurations''
by those configurations which can be transformed into 
a configuration with $A_{d+1}=\cdots = A_6=0$ 
by an appropriate SO(6) transformation.
One can then prove the following properties of 
the determinant \cite{hep-th0003223}.
For $d=5$ configurations, $\det {\cal M}$ is real. 
For $d=4$ or $d=3$ configurations, we obtain
\begin{eqnarray}
 \frac{\partial^k \Gamma}{\partial A_{\mu_1} 
\cdots \partial A_{\mu_k}} =0 \textrm{ for } k=1,\cdots, 5-d \ ,
\label{phase_effect}
\end{eqnarray}
which implies that the phase $\Gamma$ becomes more stationary 
for $d=3$ than for $d=4$.
For $d=2$ configurations, we have $\det {\cal M}=0$ and the phase $\Gamma$
becomes ill-defined.
While these properties suggest that 
the SO(6) symmetry may be broken spontaneously down to SO(3),
identifying the actual pattern of the SSB is a nontrivial issue,
which should be addressed by taking into account
the competition between the effect of the phase $\Gamma$ discussed above
and the entropic effect that favors configurations with higher dimensionality.

In order to address this issue, 
the free energy of the SO($d$) symmetric vacuum
($d=2,3,4,5$) was obtained by the GEM up to the fifth order,
and it was found that the SO$(3)$ vacuum 
has the lowest free energy \cite{1007_0883}.
This implies that the SO$(6)$ symmetry 
is spontaneously broken to SO$(3)$. 
In the SO$(d)$ vacuum,
the extent of spacetime 
$ \lambda_\mu = \frac{1}{N} \textrm{tr} (A_{\mu})^2$
in each direction
has $d$ large values
and $(6-d)$ small values, which implies that the dynamically
generated spacetime has $d$ extended directions and
$(6-d)$ shrunken directions.
This quantity $\lambda_\mu$ was calculated up to the fifth order 
in the GEM as well, and the result for $d=3$ turned out to be
\beq
\langle \lambda_\mu \rangle \simeq
\left\{
\begin{array}{ll}
1.7  & \mbox{for the three extended directions}, \\
0.2  & \mbox{for the three shrunken directions}.
\end{array}
\right.
\label{GEM_result}
\eeq
Let us also mention that
the free energy 
for the SO(2) vacuum turned out to be substantially higher
than that for the vacua with higher dimensionality \cite{1007_0883}.
This is consistent with the fact that $d=2$ configurations
are suppressed by the fermion determinant.
(See also ref.~\cite{1108_1293} for similar results in the 10D case.)


\section{Applying the 
CLM
to 
the 6D type IIB matrix model}
\label{CLM-6dIKKT}

In this section, we apply the CLM 
to the 6D version of the type IIB matrix model
(\ref{IKKT_partition})
following an analogous study on a simplified model \cite{1609_04501}.
In particular, we discuss how we probe the SSB of SO(6) symmetry
and explain the important techniques such as the gauge cooling
and the deformation, which are 
crucial in making the CLM work.

\subsection{complex Langevin equation}

Substituting (\ref{ferm_det}) in (\ref{IKKT_partition}),
we obtain
\begin{eqnarray}
Z = \int dA \, e^{-S_{\textrm{b}}} \det {\cal M} = 
\int dA \, e^{-S} \ , 
\label{IKKT_action2}
\end{eqnarray}
where the effective action 
$S = S_{\textrm{b}} - \log \det {\cal M}$
for the bosonic matrices is complex in general due to the complex
fermion determinant.
In the CLM, one considers a fictitious time evolution
of the bosonic matrices defined by the complex Langevin 
equation \cite{Parisi:1984cs,Klauder:1983sp}
\begin{eqnarray}
 \frac{d (A_{\mu})_{ij}}{dt}  = 
- \frac{\partial S}{\partial (A_{\mu})_{ji}} + 
(\eta_{\mu})_{ij}(t)   \ ,
\label{CLM_eq_1}
\end{eqnarray}
where $\eta_{\mu}(t)$ are traceless Hermitian matrices
whose elements are random variables obeying the Gaussian distribution
$\propto \exp \left( - \frac{1}{4} \int \textrm{tr} \, 
\{ \eta_\mu(t)\}^{2}  dt  \right) $.
The first term 
on the right-hand side of
eq.~(\ref{CLM_eq_1}) is called 
the drift term, which is given explicitly as
\beq
\frac{\partial S}{\partial (A_{\mu})_{ji}}
=  \frac{\partial S_{\textrm{b}}}
{\partial (A_{\mu})_{ji}} - \textrm{Tr} 
\left(  \frac{\partial {\cal M}}{\partial (A_{\mu})_{ji}} 
{\cal M}^{-1} \right)  \ ,
\label{drift-expression}
\eeq
where $\textrm{Tr}$ represents the trace 
for a $4(N^2-1) \times 4(N^2-1)$ matrix. 
The second term in (\ref{drift-expression}) is not Hermitian
as a result of the fact that the fermion determinant is complex.
Therefore, when we consider
the fictitious time evolution based on the 
complex Langevin equation (\ref{CLM_eq_1}),
we need to allow the bosonic matrices $A_{\mu}$ to evolve into
general traceless matrices.
The expectation values of an observable 
${\cal O}[A]$ with respect to the original path integral 
(\ref{IKKT_action2}) can then be calculated as
\begin{eqnarray}
 \langle {\cal O}[A] \rangle = 
\frac{1}{T} \int_{t_0} ^{t_0+T} {\cal O}[A(t)] \, dt \ ,
\label{CLM_obs}
\end{eqnarray}
where $t_0$ is the time required for thermalization, 
and $T$ should be large enough to achieve good statistics. 
Let us recall here that the observable 
${\cal O}[A]$ and the drift term (\ref{drift-expression})
are originally defined for Hermitian $A_\mu$.
In the above procedure, their definitions need to be extended
to complexified $A_\mu(t)$ by analytic continuation, which
plays a crucial role in the argument for 
justification \cite{Aarts:2009uq,Aarts:2011ax,1606_07627}.

When we solve the complex 
Langevin equation (\ref{CLM_eq_1}) numerically,
we have to discretize the fictitious time $t$.
The discretized version of (\ref{CLM_eq_1}) reads
\begin{eqnarray}
 (A_{\mu})_{ij} (t + \Delta t) = (A_{\mu})_{ij} (t) - 
\Delta t \frac{\partial S}{\partial (A_{\mu})_{ji} (t)} + 
\sqrt{ \Delta t } \, 
(\eta_{\mu})_{ij}(t) \ , 
\label{CLM_eq_2}
\end{eqnarray}
where the probability distribution of $\eta_{\mu} (t) $ is given by
$\exp \left( - \frac{1}{4} \sum_t  \textrm{tr} \, 
\{ \eta_{\mu}(t) \} ^2 \right)$.
The step-size $\Delta t$ is chosen adaptively 
at each step depending on the magnitude of the drift term \cite{0912_0617}
as described in Appendix \ref{sec:details}, where
we provide some details of our simulation.



\subsection{how to probe the SSB}
\label{sec:how-to-SSB}

As is commonly done in probing the SSB,
we introduce a term
\begin{equation}
\Delta S_{\textrm{b}} 
= \frac{1}{2}  N\varepsilon
\sum_{\mu=1}^{6} m_{\mu} \textrm{tr} (A_{\mu})^2 
\label{boson_mass} 
\end{equation}
in the action with the order $0 < m_1 \le \cdots \le m_6$, 
which breaks the SO(6) symmetry explicitly.
After taking the thermodynamic limit, which amounts to taking
the large-$N$ limit in our case,
we send the coefficient $\varepsilon$ in (\ref{boson_mass}) to zero.

As an order 
parameter for the SSB, we 
consider\footnote{Unlike the previous studies \cite{1306_6135,1509_05079}
using a method based on reweighting, we cannot use 
the eigenvalues of the tensor 
$T_{\mu \nu} =\frac{1}{N} \textrm{tr} (A_{\mu} A_{\nu})$ 
as an order parameter since 
they are not
single-valued with respect to complexified
$A_{\mu}$
and hence the relationship (\ref{CLM_obs}) does not hold.}
\begin{eqnarray}
 \lambda_{\mu} =  \frac{1}{N} 
\textrm{tr} (A_{\mu})^2 \quad \quad \mbox{~for~}\mu=1, \cdots, 6 \ , 
\label{SSB_obs_2}
\end{eqnarray}
where no sum over $\mu$ is taken. 
Note that $\lambda_\mu$ calculated for a configuration $A_{\mu}$
generated by the CLM is not necessary real because 
$A_{\mu}$ is no longer Hermitian.
However, it becomes real after taking an ensemble average 
due to the symmetry property of the drift term (\ref{drift-expression})
under $A_i \mapsto  (A_i)^\dag$ ($i=1,\cdots , 5$) and 
$A_6 \mapsto - (A_6)^\dag$.
For this reason, we take the real part of $\lambda_\mu$
when we define the expectation values $\langle \lambda_\mu \rangle$.
Due to the chosen ordering of $m_\mu$,
we have an inequality $\langle \lambda_1 \rangle \ge
\cdots \ge \langle \lambda_6 \rangle$ 
for positive $\varepsilon$. 
When we take the large-$N$ limit followed by 
the $\varepsilon \to 0$ limit, 
it can happen that 
the expectation values $\langle \lambda_{\mu} \rangle$ are not equal.
In that case we conclude that the SO$(6)$ symmetry is spontaneously broken.

Throughout this paper, we use
\begin{eqnarray}
 m_{\mu} &=& (0.5, 0.5, 1,2,4,8) \ , 
\label{SUSY_D6_mu}
\end{eqnarray}
which retains ${\rm SO}(2) \subset {\rm SO}(6)$ 
instead of breaking the SO$(6)$ symmetry completely.
The reason for making this compromise is that
having $m_1 \neq m_2$ necessarily results in 
a wider spectrum of $m_\mu$, which makes 
the $\varepsilon \rightarrow 0$ extrapolation more subtle.
This is expected not to harm anything since
it is unlikely that the SO$(6)$ symmetry is broken completely
according to the discussion
below (\ref{phase_effect}).


\subsection{gauge cooling for the excursion problem}

In order to justify the CLM, the probability distribution of the 
magnitude of the drift term
\begin{eqnarray}
 u = \sqrt{\frac{1}{6N^3} \sum_{\mu=1}^{6} 
\sum_{i,j=1}^{N} \left|  
\frac{\partial S}{\partial (A_{\mu})_{ji}} \right|^2 }
\label{drift_norm_def}
\end{eqnarray}
should fall off exponentially or faster \cite{1606_07627}. 
There are two sources for the violation of this property.
One is the ``excursion problem'', which occurs 
when $A_{\mu}$ develops a large anti-Hermitian part.
The other is the ``singular-drift problem'', 
which occurs 
because of the appearance of ${\cal M}^{-1}$ in (\ref{drift-expression})
when 
some eigenvalues of ${\cal M}$ come close to zero frequently.

In order to avoid the excursion problem, we use a technique 
called gauge cooling \cite{Seiler:2012wz}, 
which keeps $A_{\mu}$ as close to Hermitian matrices as possible. 
This amounts to minimizing the ``Hermiticity norm'' defined by
\begin{eqnarray}
 {\cal N}_{\textrm{H}} = - \frac{1}{6N} 
\sum_{\mu=1}^6 \textrm{tr} \{ (A_{\mu} - A_{\mu}^{\dagger})^2 \}
\label{hermiticity_norm}
\end{eqnarray}
by performing an SL$(N, {\bf C})$ transformation
$A_{\mu} \to g A_{\mu} g^{-1}$, where
\begin{eqnarray}
g = e^{- \alpha G} \ , \quad \quad
G = \frac{1}{N} \sum_{\mu=1}^{6} [A_{\mu}, A_{\mu}^{\dagger}]  \ ,
\label{gauge_cooling}
\end{eqnarray}
after each step of solving 
the discretized Langevin equation (\ref{CLM_eq_2}).
Here, $G$ is the gradient of $\displaystyle {\cal N}_{\textrm{H}}$ 
with respect to the SL$(N, {\bf C})$ transformation, which is derived
in ref.~\cite{1609_04501}.
We choose the real positive parameter $\alpha$ 
in such a way that the Hermiticity norm $\displaystyle {\cal N}_{\textrm{H}}$ 
is minimized.
In refs.~\cite{Nagata:2015uga,1606_07627},
it was proven
that adding the gauge cooling procedure in the CLM does not
affect the argument for its justification.

\subsection{deformation for the singular-drift problem}
\label{sec:deform-singular-drift}

In order to avoid the singular-drift problem, 
we deform the fermionic action by adding
\begin{eqnarray}
 \Delta S_{\textrm{f}} &=& N m_{\rm f} \,
 \textrm{tr} ({\bar \psi}_{\alpha} 
(\Gamma_6)_{\alpha \beta} \psi_{\beta}) \ ,
\label{ferm_mass}
\end{eqnarray}
where $m_{\rm f}$ is the deformation parameter.
This technique was proposed originally in ref.~\cite{1609_04501},
where the singular-drift problem was indeed overcome in
an SO(4) symmetric matrix model with a complex fermion determinant.
The fermionic mass term (\ref{ferm_mass}) modifies 
the linear transformation ${\cal M}$ in (\ref{defM}) 
to  $\widetilde{\cal M}$ given by
\beq
\Psi_\alpha  \mapsto (\widetilde{\cal M} \Psi)_\alpha \equiv 
 (\Gamma_{\mu})_{\alpha \beta} [A_{\mu}, \Psi_{\beta}] 
+ m_{\rm f} \Psi_\alpha 
\label{defMtilde}
\eeq
using $\Gamma_6  = {\bf 1} \otimes {\bf 1}$,
which implies that 
the eigenvalue distribution of the matrix ${\cal M}$ 
is shifted in the direction of the real axis for the same $A_\mu$.
We will see that this effect enables us to avoid the eigenvalues
coming close to zero.

Note that we cannot construct an SO(6) symmetric mass term
for $\psi_{\alpha}$
since $\psi_{\alpha}$ transforms as a Weyl spinor under
an SO(6) transformation.
The term (\ref{ferm_mass}) we have chosen is a kind of mass term,
which breaks the SO(6) symmetry minimally to SO(5).
We first investigate whether this SO(5) symmetry of the deformed model
is spontaneously broken for various values of $m_{\rm f}$,
and then discuss what happens as $m_{\rm f}$ is decreased.

Let us also note that the $m_{\rm f}\rightarrow \infty$ limit
of the deformed model is nothing but
the SO(6) symmetric bosonic model
since fermionic matrices $\psi_\alpha$ obviously decouple in that limit.
Thus, the deformation (\ref{ferm_mass}) may be regarded
as an interpolation between the dimensionally reduced super Yang-Mills model
and its bosonic counterpart, in which the fermionic matrices are 
omitted.
It is known that the SO($D$) symmetry of the $D$-dimensional version 
of the bosonic type IIB matrix model is not 
spontaneously broken \cite{hep-th9811220}.
We use this fact to test the validity of our method.

As the deformation paramter $m_{\textrm{f}}$ is decreased, the effects
of fermionic matrices are gradually turned on.
From the trend for decreasing $m_{\textrm{f}}$, we try to draw some
conclusions on the undeformed model, assuming that nothing dramatic
occurs in the vicinity of $m_{\textrm{f}}=0$.
This strategy has been quite successful 
in the simplified model \cite{1609_04501},
where the spontaneous breaking of SO(4) symmetry to SO(2) was
confirmed as predicted by the GEM \cite{Nishimura:2004ts}.

To summarize, the model we investigate by the CLM is defined by 
the partition function 
(\ref{IKKT_action2}), where
$S_{\textrm{b}}$ and ${\cal M}$ are replaced with
$\tilde{S}_{\textrm{b}} = S_{\textrm{b}} + \Delta S_{\textrm{b}}$ and 
$\widetilde{\cal M}$, respectively.
The drift term
(\ref{drift-expression}) in 
the complex Langevin equation (\ref{CLM_eq_1})
should be modified accordingly.

\begin{figure}[htbp]
\centering{}
\includegraphics[width=8.0cm]{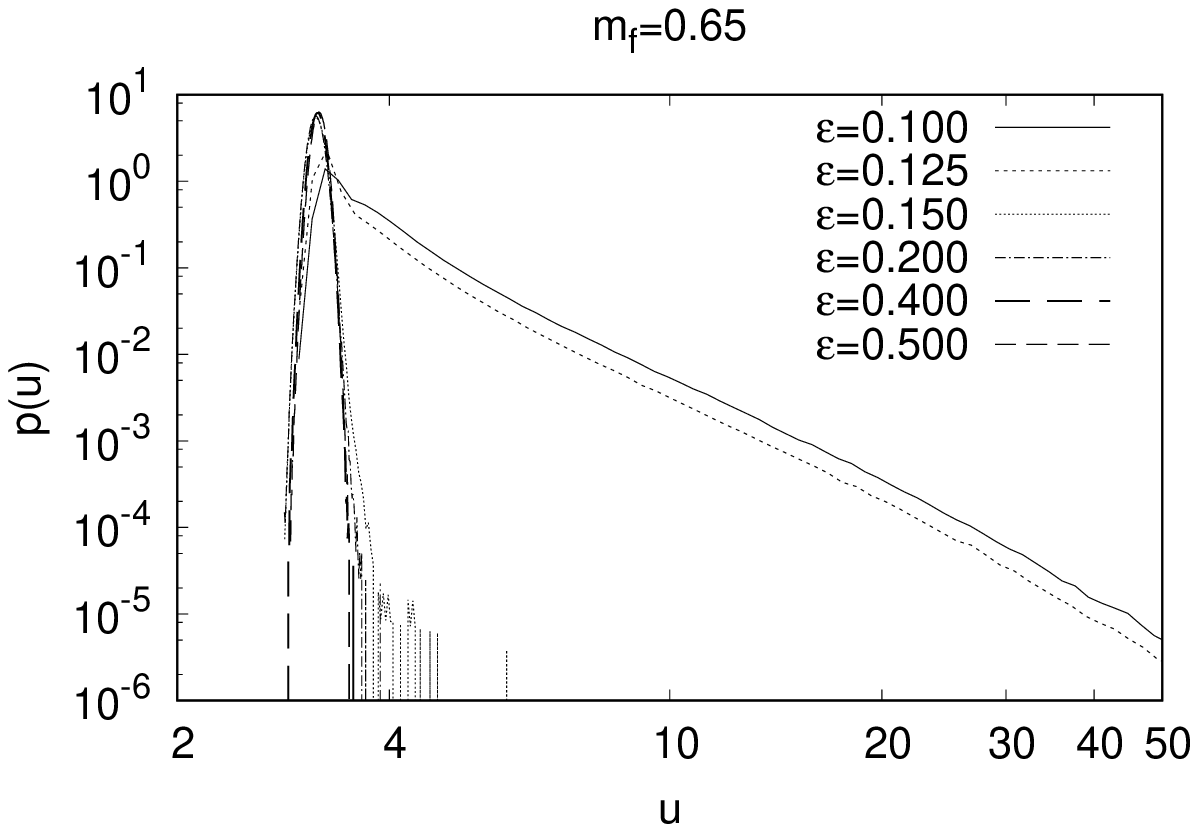}
\includegraphics[width=8.0cm]{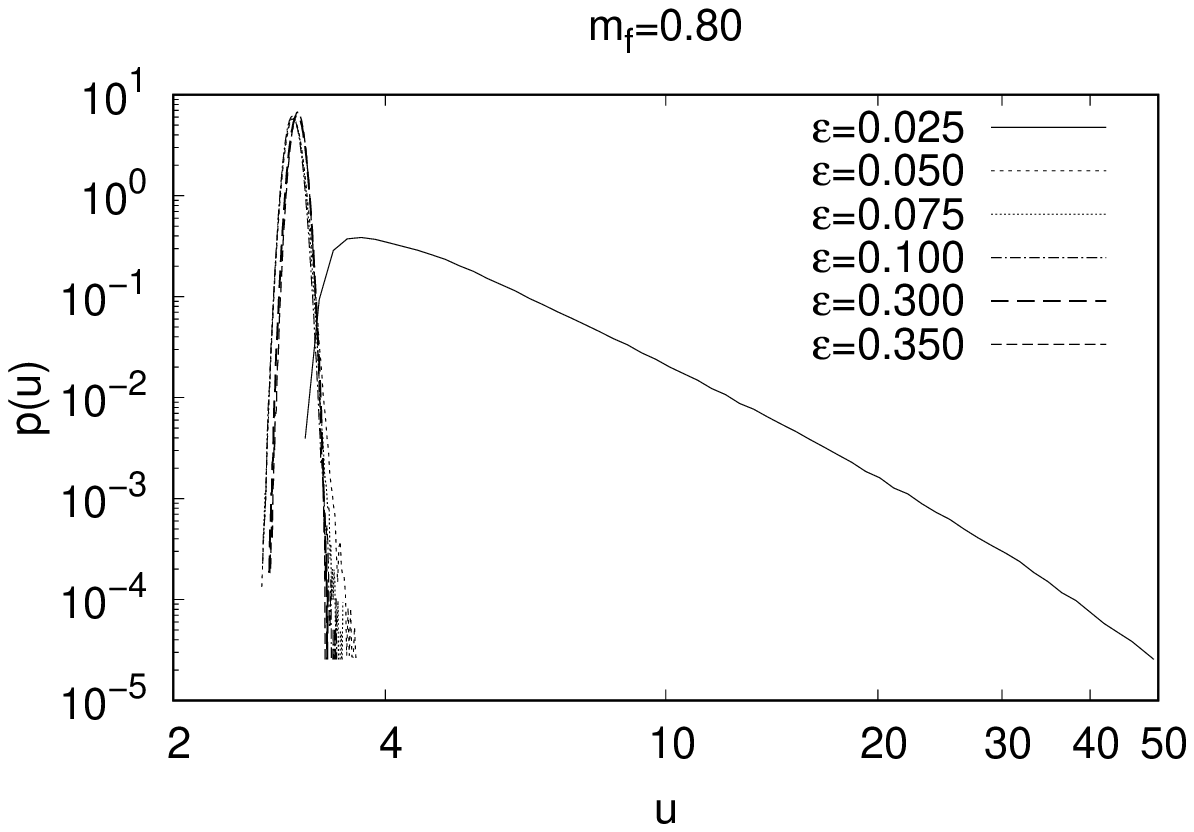}
\includegraphics[width=8.0cm]{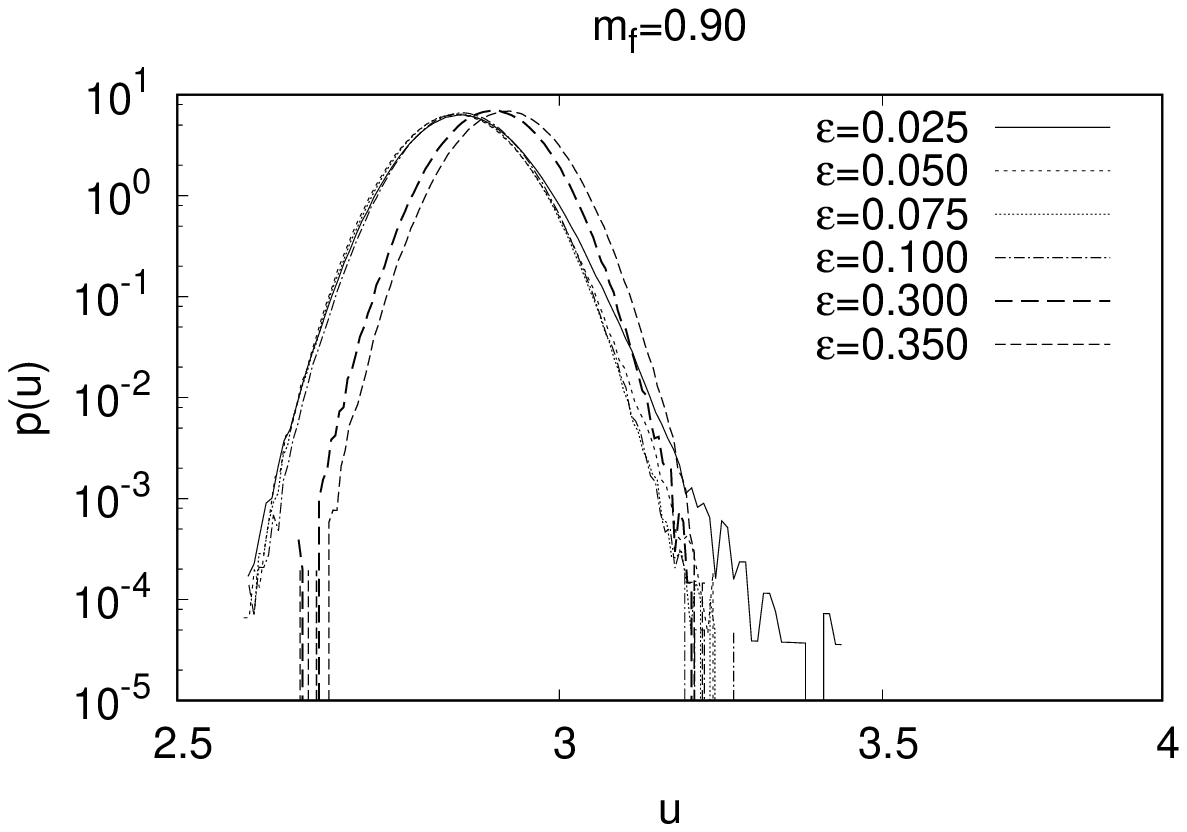}
\caption{The probability distribution $p(u)$ of the magnitude of the
drift term $u$ defined by eq.~(\ref{drift_norm_def}) 
is plotted for $N=24$ with $ m_{\textrm{f}} = 0.65$ (Top-Left), 
$ m_{\textrm{f}} = 0.80$ (Top-Right) 
and $ m_{\textrm{f}} = 0.90$ (Bottom).}
\label{histlog_dnorm_ex}
\end{figure}

While the singular-drift problem is cured for large enough $m_{\rm f}$,
the problem occurs for moderate $m_{\rm f}$
with $\varepsilon$ smaller than some value depending on $m_{\rm f}$.
Therefore, when we make an $\varepsilon\rightarrow 0$ extrapolation,
we have to choose carefully the data points that do not suffer from 
this problem.
For that purpose, we probe
the probability distribution $p(u)$ of 
the magnitude of the drift term $u$
defined by eq.~(\ref{drift_norm_def}). 
In Fig.~\ref{histlog_dnorm_ex}
we show the log-log plot of $p(u)$ 
for $N=24$ with $ m_{\textrm{f}} = 0.65, 0.8, 0.9$. 
We find at $m_{\textrm{f}} = 0.65$ that $p(u)$ falls off
exponentially or faster for $ \varepsilon \geq 0.15$, 
while a clear power-law tail 
develops
for
$ \varepsilon \leq 0.125$.
Therefore, we can trust only the results for
$ \varepsilon \geq 0.15$ according to the 
criterion of ref.~\cite{1606_07627}.
Similarly, at $m_{\textrm{f}} = 0.80$,
we find that $p(u)$ falls off exponentially or faster 
for $ \varepsilon \geq 0.05$.
At $m_{\textrm{f}} = 0.90$, we observe such a behavior
for all values of $\varepsilon$ investigated.

\begin{figure}[htbp]
\centering{}
\includegraphics[width=8.0cm]{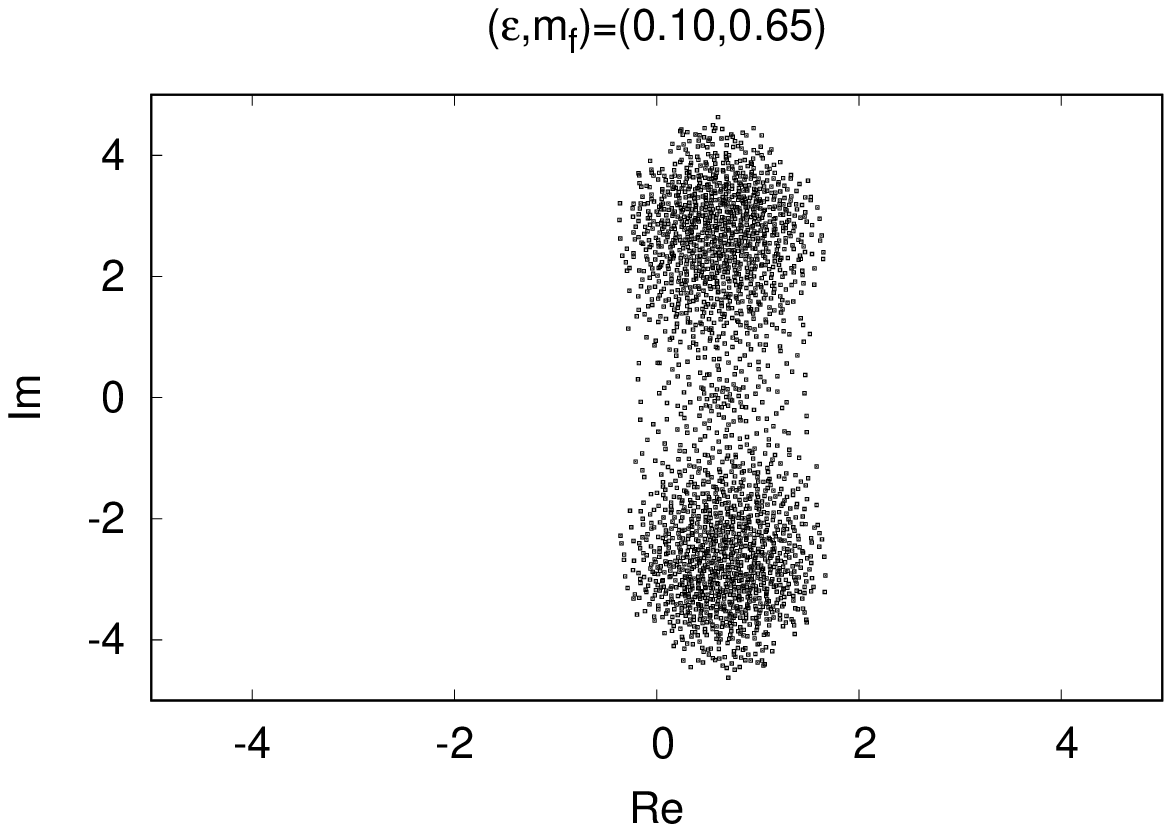}
\includegraphics[width=8.0cm]{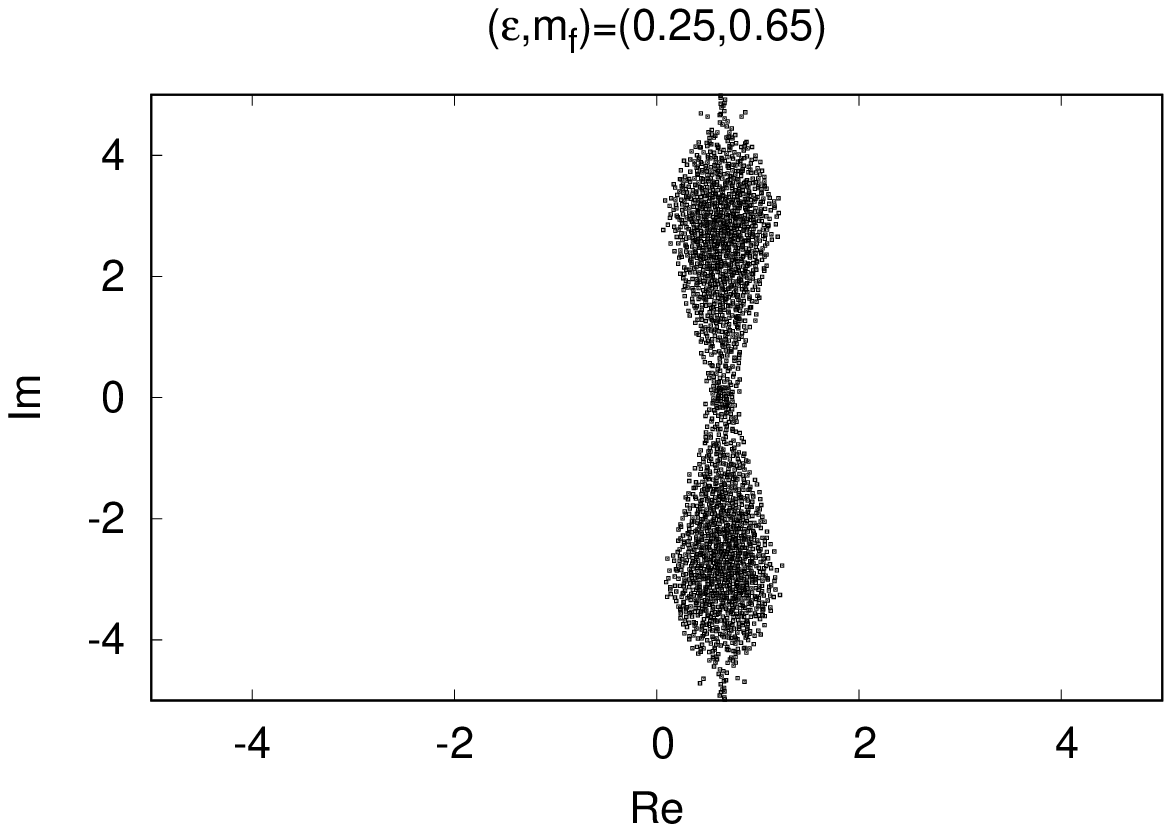}
\caption{The scatter plot 
of the eigenvalues of the $4(N^2-1) \times 4 (N^2-1)$ matrix 
$\widetilde{\cal M}$ 
obtained with a thermalized configuration for 
$N=24$ and $m_{\textrm{f}} = 0.65$
with $ \varepsilon = 0.1$ (Left) and $ \varepsilon = 0.25$ (Right).}
\label{scatter_m_eig}
\end{figure}

In order to understand how the singular-drift problem is avoided
by the deformation (\ref{ferm_mass}),
we show in Fig.~\ref{scatter_m_eig}
the scatter plot of the eigenvalues of 
the $4(N^2-1) \times 4 (N^2-1)$ matrix $\widetilde{\cal M}$ 
obtained with
a thermalized configuration
for $N=24$ and $m_{\textrm{f}} = 0.65$
with $ \varepsilon = 0.1$ (Left) and $ \varepsilon = 0.25$ (Right).
Note that the eigenvalue distribution is shifted in the direction
of the real axis compared with that for $m_{\rm f}=0$ 
as one can deduce from (\ref{defMtilde}).
From this figure, we find that eigenvalues close to zero appear
for $\varepsilon = 0.1$, but not for $\varepsilon = 0.25$.
%

\section{Results} 
\label{sec:result}

In this section we present our results obtained in the way described
in the previous section.
Let us recall that the deformation 
(\ref{ferm_mass})
breaks the SO(6) symmetry to SO(5).
In order to probe the SSB of SO(5) that remains for $m_{\rm f} \neq 0$, 
we need to take 
the $N \rightarrow  \infty$ limit first, and then
the $\varepsilon \to 0$ limit.
Finally, in order to obtain results for the undeformed model,
we need to extrapolate $m_{\rm f}$ to zero.

\begin{figure}[htbp]
\centering{}
\includegraphics[width=8.0cm]{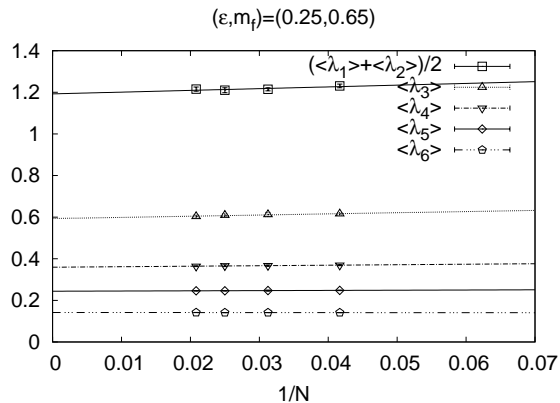}
\caption{The expectation values
$\displaystyle \langle \lambda_{\mu} \rangle$ are plotted against
$1/N$ for 
$\varepsilon = 0.25$ and $ m_{\textrm{f}}=0.65$ with $N=24,32,40,48$.
For $\langle \lambda_1 \rangle$ and $\langle \lambda_2 \rangle$,
we plot their average to increase the statistics.
The straight lines represent fits to $a + b/N$,
which enable us to make reliable extrapolations to $N=\infty$.
}
\label{large_N_ex}
\end{figure}

First we discuss how we take the $N \rightarrow  \infty$ limit.
In Fig.~\ref{large_N_ex}
we plot the expectation values 
$\displaystyle \langle \lambda_{\mu} \rangle$
for 
$\varepsilon = 0.25$ and 
$m_{\textrm{f}} = 0.65$ with $N=24,32,40,48$ against $1/N$.
For $\langle \lambda_1 \rangle$ and $\langle \lambda_2 \rangle$,
we plot the average to increase the statistics
since we know theoretically that they should be equal due to 
the chosen parameters (\ref{SUSY_D6_mu}) for the symmetry breaking term.
We find that our data can be fitted nicely to straight lines,
which implies that our data follow
the large-$N$ asymptotic behavior $a + b/N$. Similar behaviors were observed
for other $\displaystyle (\varepsilon, m_{\textrm{f}})$ as well. 
Thus we can make extrapolations to $N=\infty$ reliably.
Let us denote the extrapolated values obtained for each 
$\displaystyle (\varepsilon, m_{\textrm{f}})$
as
$\displaystyle \langle \lambda_{\mu} \rangle_{\varepsilon, m_{\textrm{f}}}$.

\begin{figure}[htbp]
\centering{}
\includegraphics[width=8.0cm]{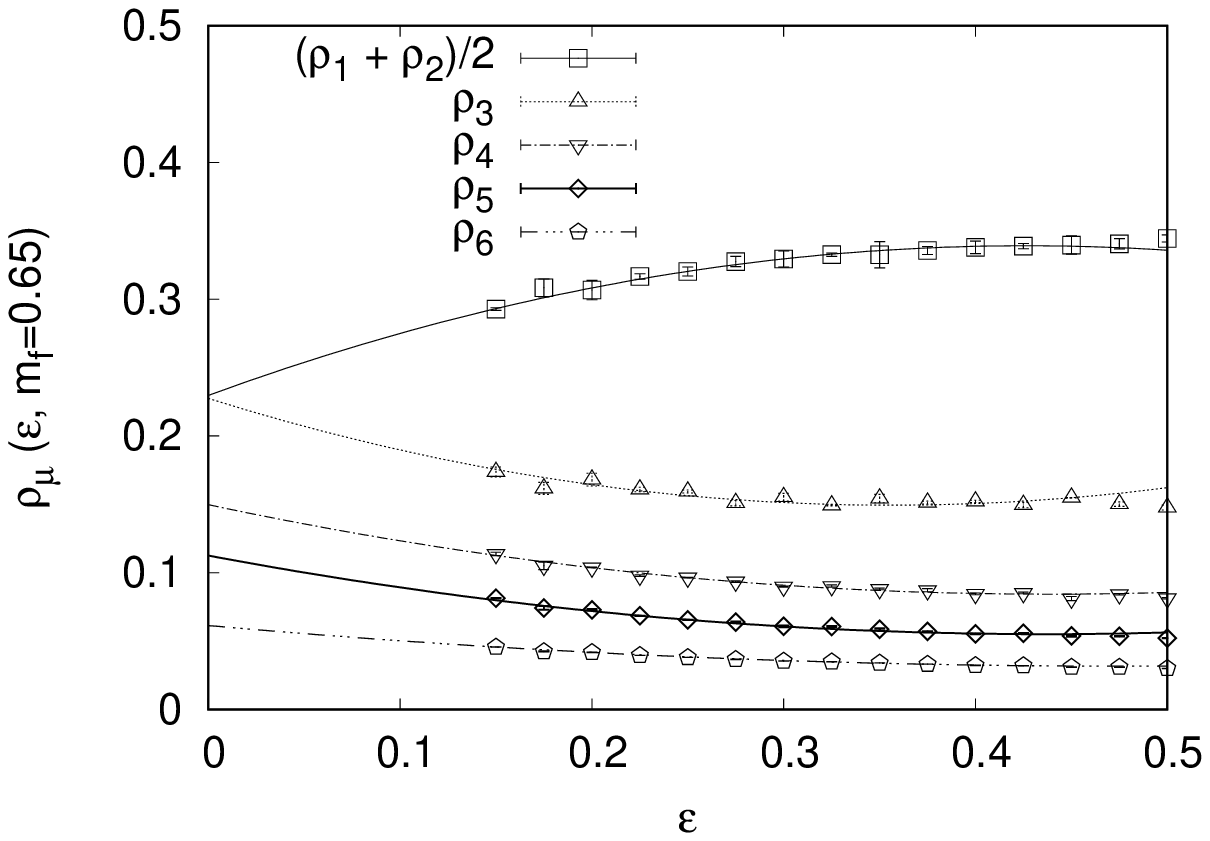}
\includegraphics[width=8.0cm]{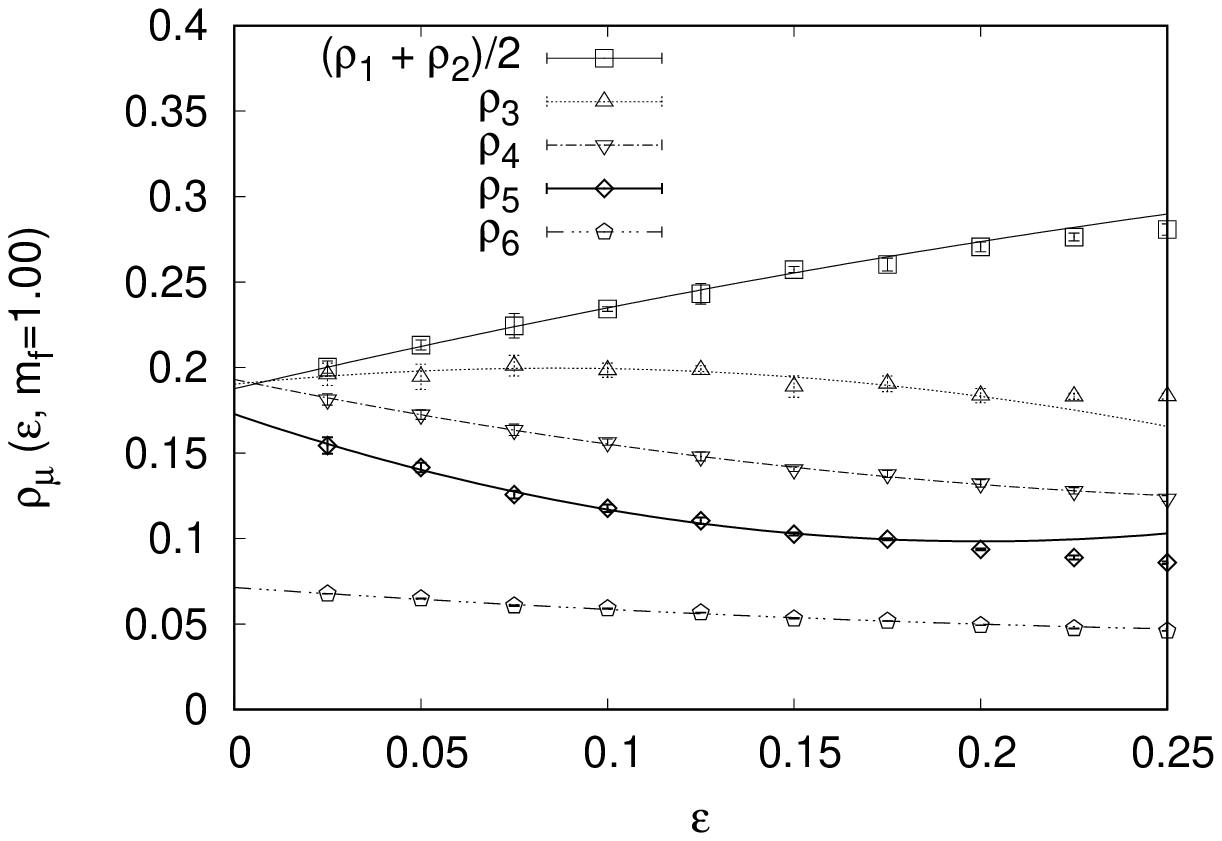}
\includegraphics[width=8.0cm]{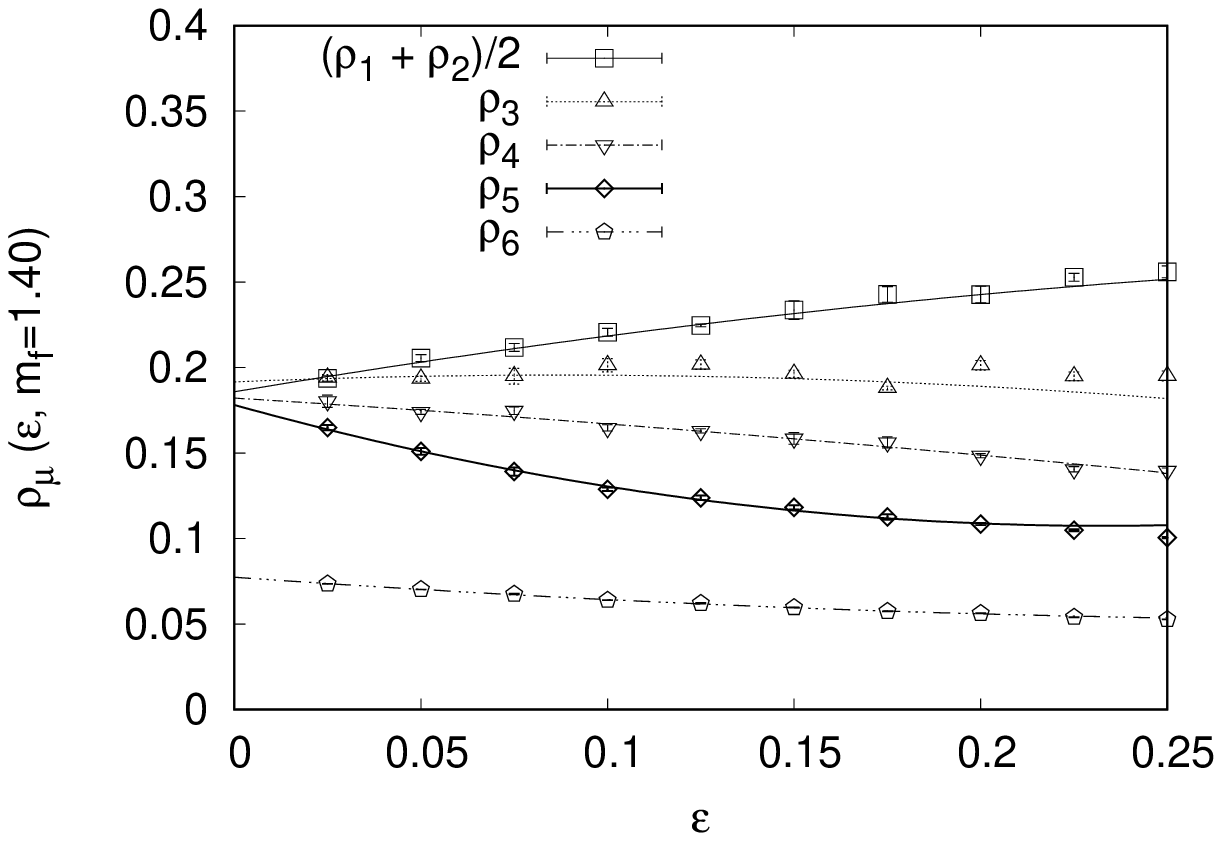}
\includegraphics[width=8.0cm]{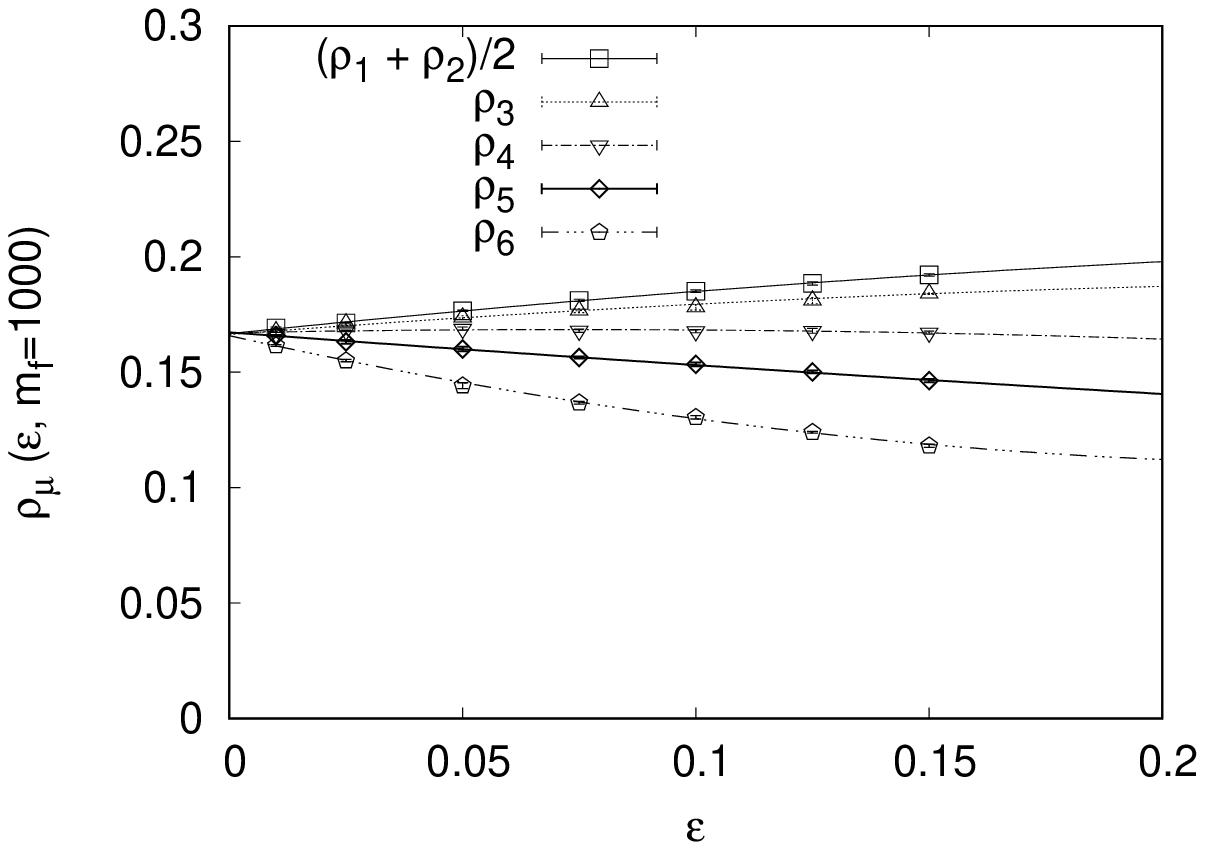}
\caption{The ratios
$\rho_{\mu} (\varepsilon, m_{\textrm{f}})$ 
defined by (\ref{rho_ratio}) are plotted
against $\varepsilon$ for the deformed model
with $m_{\textrm{f}} = 0.65$ (Top-Left), 
$m_{\textrm{f}} = 1.0$ (Top-Right), 
$m_{\textrm{f}} = 1.4$ (Bottom-Left)
and 
$m_{\textrm{f}} = 1000$ (Bottom-Right).
An average is taken for $\rho_1$ and $\rho_2$
according to the treatment in Fig.~\ref{large_N_ex}.
The lines represent fits to the quadratic form 
$a + b \varepsilon + c { \varepsilon }^2$. 
}
\label{e-fit}
\end{figure}

Next we make an extrapolation to 
$ \varepsilon = 0$. 
For that purpose, it is convenient to consider the ratio \cite{1609_04501}
\begin{eqnarray}
 \rho_{\mu} (\varepsilon, m_{\textrm{f}}) = 
\frac{\langle \lambda_{\mu} \rangle_{\varepsilon, m_{\textrm{f}}}}
{\sum_{\nu=1}^{6} \langle \lambda_{\nu} \rangle_{\varepsilon, m_{\textrm{f}}}} \ ,
\label{rho_ratio}
\end{eqnarray}
in which a large part of the $\varepsilon$ dependence cancels between
the numerator and the denominator thereby making 
the $\varepsilon \rightarrow 0 $ extrapolation more reliable.
In Fig.~\ref{e-fit}, we plot 
$\rho_{\mu} (\varepsilon, m_{\textrm{f}})$
against $\displaystyle \varepsilon$ for 
$m_{\textrm{f}} = 0.65$, 1.0, 1.4 and 1000. 
Here we plot only the data that do not suffer from the singular-drift problem
based on the criterion explained in section \ref{sec:deform-singular-drift}.
We make an $\displaystyle \varepsilon \to 0$ extrapolation
by fitting our data to the quadratic form\footnote{We have also tried
including a higher order term 
$\propto {\varepsilon }^3$ in the fitting function,
but the extrapolated values were unaffected within the fitting error.} 
$a + b \varepsilon + c { \varepsilon }^2$ 
with the fitting range
$ 0.15 \leq \varepsilon \leq 0.475$ for $ m_{\textrm{f}} = 0.65 $,
$ 0.025 \leq \varepsilon \leq 0.175$ for $ m_{\textrm{f}} = 1.0$,
$0.025 \leq \varepsilon \leq 0.2$ for $m_{\textrm{f}} = 1.4$
and 
$0.01 \leq \varepsilon \leq 0.15$ for $m_{\textrm{f}} = 1000$.
At $ m_{\textrm{f}} = 0.65$, 
the fitting curves for $(\rho_1 + \rho_2)/2$ and $\rho_3$
are seen to merge at $\varepsilon=0$, which implies that
the SO(5) symmetry of the deformed model 
is spontaneously broken to SO(3).
Similarly, at $ m_{\textrm{f}} = 1.0$
we observe an SSB from SO(5) to SO(4),
whereas at $ m_{\textrm{f}} = 1.4$, 
we find that
the SO(5) symmetry of the deformed model is not broken spontaneously.

Let us recall that in the large-$m_{\textrm{f}}$ limit,
the deformed model reduces to the SO(6) symmetric 
bosonic model, in which the SSB of SO(6) is known not to
occur \cite{hep-th9811220}.
Indeed our results for $ m_{\textrm{f}} = 1000$
in Fig.~\ref{e-fit} (Bottom-Right) show that
the fitting curves for $(\rho_1 + \rho_2)/2$ and 
$\rho_\mu$ ($\mu=3, \cdots ,6$)
all merge at $\varepsilon=0$ as expected.
This confirms that the SSB observed for smaller $m_{\textrm{f}}$
is a physical effect, which cannot be attributed to some artifacts 
in the $\varepsilon \rightarrow 0$ extrapolations.

\begin{figure}[htbp]
\centering{}
\includegraphics[width=8.0cm]{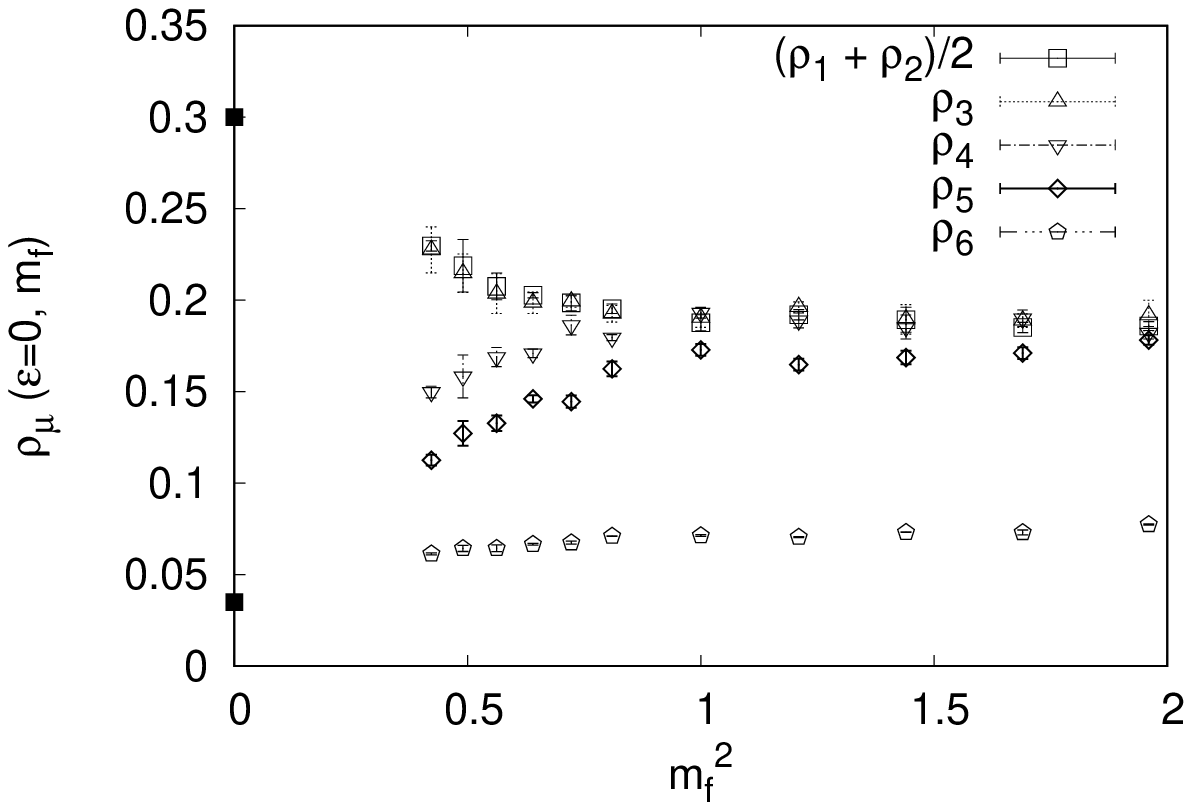}
\includegraphics[width=8.0cm]{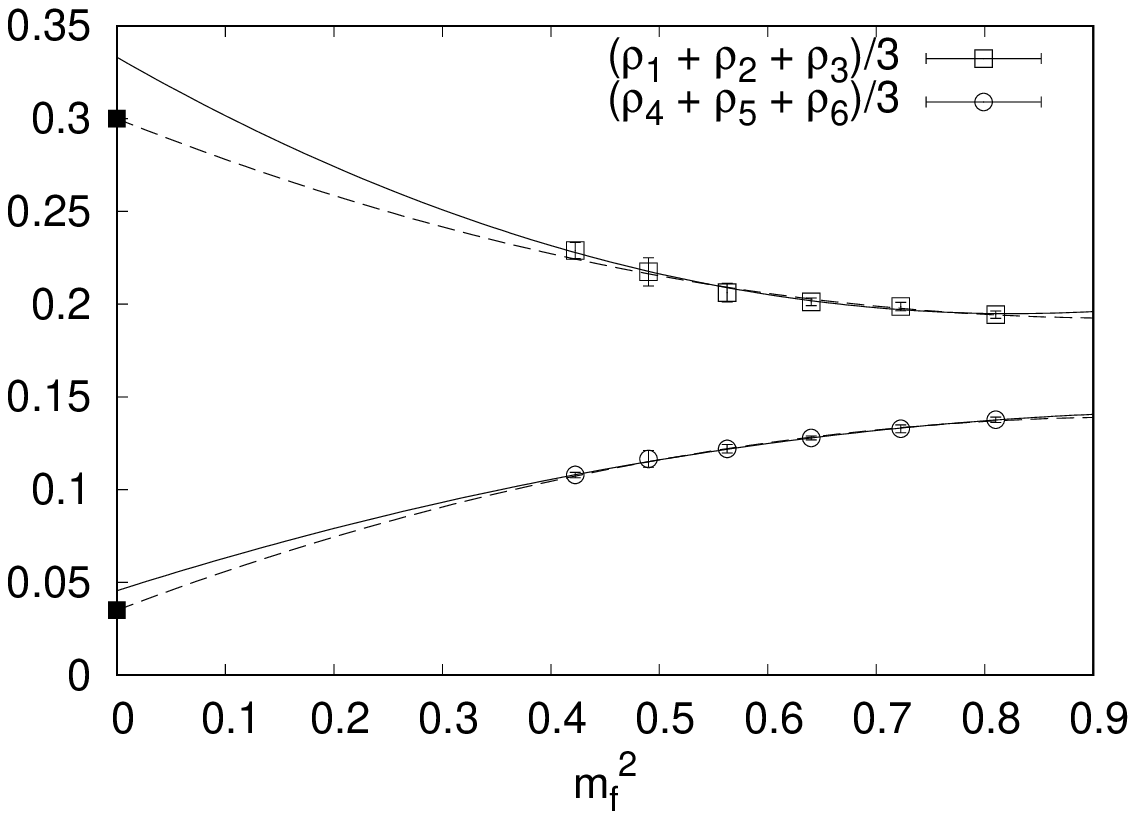}
\caption{(Left) The $\varepsilon \to 0$ extrapolated values
of the ratios
$\rho_{\mu} (\varepsilon, m_{\textrm{f}})$
defined by (\ref{rho_ratio}) are plotted
against ${m_{\textrm{f}}}^2$. 
The filled squares at $m_{\textrm{f}}=0$
represent the GEM prediction (\ref{GEM_result2})
for $(\rho_1,\rho_2, \rho_3)$ and $(\rho_4 , \rho_5, \rho_6)$, respectively.
(Right) The averages $(\rho_1+\rho_1+\rho_3)/3$
and $(\rho_4+\rho_5+\rho_6)/3$ of the ratios
are plotted against ${m_{\textrm{f}}}^2$. 
The solid lines represent fits to 
$\displaystyle a + b {m_{\textrm{f}}}^2 + c { m_{\textrm{f}} }^4$,
whereas the dashed lines represent fits to the same form 
constrained to pass through the
points at $m_{\rm f}=0$ predicted from the GEM.
}
\label{mf-fit}
\end{figure}

In Fig.~\ref{mf-fit} (Left), 
we plot the $\varepsilon \rightarrow 0$ extrapolated values
of $\rho_{\mu} (\varepsilon, m_{\textrm{f}})$ 
against ${m_{\textrm{f}}}^2$ for 
$m_{\textrm{f}} = 0.65$,
0.7, 0.75, 0.8, 0.85, 0.9, 1.0, 1.1, 1.2, 1.3, 1.4. 
We find that an SO(3) vacuum is realized
for $m_{\textrm{f}} \lesssim  0.9$, 
while an SO(4) vacuum is realized
for $1.0 \lesssim  m_{\textrm{f}} \lesssim  1.3$.
Judging from the pattern of the SSB with decreasing $m_{\rm f}$, 
it is reasonable to 
consider that the symmetry that survives for
$m_{\rm f} < 0.65$ is \emph{at most} SO(3).
Note also that an SO(2) vacuum is suppressed by the fermion determinant
as we discussed below (\ref{phase_effect}).
Combining this argument with our results, we conclude that
the SO(6) rotational symmetry is spontaneously broken to SO(3)
in the undeformed model corresponding to $m_{\rm f}=0$,
which agrees with the prediction from the GEM.

Note that the results (\ref{GEM_result}) obtained by 
the GEM \cite{1007_0883} can be rephrased 
in terms of the ratios $\rho_\mu$ as
\beq
 \rho_1 = \rho_2 = \rho_3 \simeq \frac{1.7}{5.7} \simeq 0.3 \ , \quad \quad 
\rho_4  \simeq \rho_5  \simeq \rho_6  \simeq \frac{0.2}{5.7} \simeq 0.035 \ . 
\label{GEM_result2}
\eeq
These values are represented by the 
filled squares at $m_{\textrm{f}}=0$ in Fig.~\ref{mf-fit}.
As $m_{\rm f}$ decreases,
the data points for $(\rho_1+\rho_2)/2$
and $\rho_3$ are coming closer to 0.3,
whereas the data points for $\rho_4$, $\rho_5$ and $\rho_6$ are
coming closer to 0.035.
For the sake of clearer comparison,
we plot in Fig.~\ref{mf-fit} (Right), 
the averages $(\rho_1+\rho_2+\rho_3)/3$
and $(\rho_4+\rho_5+\rho_6)/3$
for $m_{\rm f} \le 0.9$ corresponding to the SO(3) symmetric phase.
The asymptotic behavior of these quantities for 
$m_{\textrm{f}} \to 0$ is expected to be
a power series with respect to ${m_{\textrm{f}}}^2$
due to the symmetry under
$\displaystyle m_{\textrm{f}} \to -m_{\textrm{f}}$.
We therefore fit our data points to the form
$\displaystyle a + b {m_{\textrm{f}}}^2 + c { m_{\textrm{f}} }^4$
with the fitting range $ 0.65 \leq {m_{\textrm{f}}} \leq 0.9$,
and the extrapolation to $m_{\rm f}=0$ yields
\begin{eqnarray}
\frac{\rho_1+\rho_2+\rho_3}{3}  = 0.33(2) \ , \quad \quad
\frac{\rho_4+\rho_5+\rho_6}{3}  = 0.046(3) \ , 
\label{final_result}
\end{eqnarray}
which are close to the values (\ref{GEM_result2})
predicted by the GEM.
In fact, we can make reasonable fits
passing through the points at $m_{\rm f}=0$
predicted from the GEM as shown by the dashed lines.
Note, however, that the GEM involves a truncation,
which necessarily yields certain systematic errors.
Hence, precise agreement is not anticipated.

\section{Summary and discussions}
\label{sec:summary}

In this paper we have discussed the SSB of rotational symmetry
conjectured to occur in dimensionally reduced super Yang-Mills models
for $D=6$ and $D=10$.
In particular, the $D=10$ case is relevant to 
the dynamical generation of four-dimensional space-time 
in a nonperturbative formulation of superstring theory in ten dimensions.
It is known that the phase of the complex fermion determinant
should play a crucial role,
which implies that a first principle investigation of this issue 
based on Monte Carlo methods necessarily faces a severe sign problem.

Extending the previous work \cite{1609_04501} on a simplified model, 
we have investigated the $D=6$ case using the CLM.
In particular, we use the deformation technique to overcome the
singular-drift problem, which occurs in many interesting systems
with a complex fermion determinant including finite density QCD at low
temperature.
In the data analysis, it is important to probe the probability distribution
of the drift term so that we can determine 
the parameter region in which the CLM is valid.
After taking the large-$N$ limit,
we are able to observe that the remaining SO(5) symmetry
of the deformed model is spontaneously broken to
SO(4) or to SO(3) as the deformation parameter $m_{\rm f}$ is decreased.
Combining this result with 
the argument that an SO(2) vacuum is suppressed by the 
fermion determinant, we 
conclude that an SO(3) vacuum is chosen
in the undeformed model, which is consistent with the prediction of the GEM.
Extrapolations to $m_{\rm f}=0$ for the extent of spacetime in each direction
also give results consistent with the values predicted by the GEM.
These results 
go far beyond those obtained by
a reweighting-based method \cite{1306_6135}.

An application of the same method to the $D=10$ case is 
ongoing \cite{D_10_future}.
In this case the GEM results \cite{1108_1293}
suggest that the SO(10) symmetry is spontaneously broken to SO(3)
as opposed to the original expectation that 
the SO(4) symmetry survives.
Whether we can observe an SO(3) vacuum for nonzero $m_{\rm f}$ 
already gives us an important clue on this issue.
The fact that the CLM with the deformation technique
turned out to be useful in investigating such 
a physically interesting issue that has been 
hard to address
due to the severe sign problem 
encourages us to apply it also to
other interesting complex action systems 
such as
finite density 
QCD \cite{Nagata:2017pgc}.

\section*{Acknowledgements}
We thank H.~Kawai and H.~Steinacker for valuable comments and discussions. 
We are also grateful to L.L.~Salcedo for his comment on footnote 8,
which is reflected in this revised version.
This research was supported by MEXT as ``Priority Issue on Post-K computer''
(Elucidation of the Fundamental Laws and Evolution of the Universe) 
and Joint Institute for Computational Fundamental Science (JICFuS). 
Computations were carried out using computational resources such as 
KEKCC, NTUA het clusters and FX10 at Kyushu University. 
T.~A.\ was supported in part by Grant-in-Aid  for Scientific Research 
(No.~17K05425) from Japan Society for the Promotion of Science.

\appendix
\section{Details of the complex Langevin simulation}
\label{sec:details}

In this section we provide some details of our complex 
Langevin simulation.

First we discuss 
the idea of adaptive step-size \cite{0912_0617}
used in solving the discretized complex Langevin 
equation (\ref{CLM_eq_2}) numerically.
The point here is that the drift term in (\ref{CLM_eq_2}) 
can become large occasionally, and the associated 
discretization artifacts can make the simulation imprecise
or even unstable in the worse case.
In order to avoid these problems, we probe the magnitude of the 
drift term (\ref{drift_norm_def}) at each step, and when it
gets larger than a certain threshold value $u_0$, we decrease the
step-size as $\Delta t = (\Delta t)_0 \times u_0/u$ (for $u\ge u_0$).
A fixed step-size $(\Delta t)_0 = 10^{-5}$ is used during the
thermalization process, and
the threshold value $u_0$ is determined by taking an average of 
(\ref{drift_norm_def}) during that process.

Next we discuss how we estimate 
the second term in the drift term (\ref{drift-expression}),
which is the most time-consuming part of our calculation.
Since the matrix ${\cal M}$ has the size $4(N^2-1)\times 4(N^2-1)$,
direct calculation of ${\cal M}^{-1}$ would require
${\rm O}(N^6)$ arithmetic operations.
We can reduce this cost to ${\rm O}(N^3)$
by using
the so-called noisy estimator.\footnote{This technique
was not used in the previous work on a simplified model \cite{1609_04501},
in which the fermionic variables were in the fundamental representation 
of SU($N$),
and hence the calculation of the corresponding drift term required
only O$(N^3)$ arithmetic operations even with the direct method.}


The idea is based on the identity
\begin{eqnarray}
 \textrm{Tr} \left(  \frac{\partial {\cal M}}{\partial (A_{\mu})_{ji}} 
{\cal M}^{-1} \right) =
\left\langle 
\chi^{*} 
\frac{\partial {\cal M}}{\partial (A_{\mu})_{ji}} {\cal M}^{-1} 
\chi 
\right\rangle_{\chi} 
\ , \label{noisy_idea2} 
\end{eqnarray}
where the average $\langle \ \cdot \ \rangle_{\chi} $ is taken
with respect to the Gaussian noise $\chi$,
which represents a $4(N^2-1)$-dimensional vector whose
elements are complex Gaussian variables normalized as
$\langle \chi ^{*}_k \chi_l \rangle = \delta_{kl}$.
In practice, we generate the Gaussian noise only once and
estimate the trace using it
instead of taking the average on the right hand side of 
(\ref{noisy_idea2}).
The use of this approximation does not yield any 
systematic errors in the CLM in the $\Delta t \rightarrow 0$ 
limit since the associated Fokker-Planck equation 
remains the same \cite{Batrouni:1985jn}.

The quantity $\zeta = {\cal M}^{-1} \chi$ in (\ref{noisy_idea2})
can be calculated by solving the linear equation
\begin{eqnarray}
 {\cal M}^{\dagger} {\cal M} \zeta = {\cal M}^{\dagger} \chi  \ ,
\label{noisy_idea3}
\end{eqnarray}
where ${\cal M}^{\dagger} {\cal M}$ is a Hermitian 
positive semi-definite matrix,
using the conjugate gradient method.
This method is an iterative one, in which
one acts ${\cal M}$ and ${\cal M}^\dagger$
on a $4(N^2-1)$-dimensional vector many times.
If one does this directly using an explicit representation of 
${\cal M}$, it requires ${\rm O}(N^4)$ arithmetic operations.
In fact,
one can perform this calculation
using (\ref{defM}) for ${\cal M}$ and
\beq
\Psi_\alpha  \mapsto ({\cal M}^\dagger \Psi)_\alpha \equiv 
 (\Gamma_{\mu}^\dagger)_{\alpha \beta} [A_{\mu}^\dagger , \Psi_{\beta}]
\label{defMdag}
\eeq
for ${\cal M}^\dagger$, which requires only
${\rm O}(N^3)$ arithmetic operations.
In the parameter region in which the CLM is valid,
the number of iterations required for convergence
of the conjugate gradient method 
is almost independent of $N$, and 
it is typically of the order of $100$.
Thus the computational cost of our simulation grows
only as O$(N^3)$ with the matrix size $N$.



\end{document}